\documentclass[11pt,a4paper]{article}

\usepackage{graphicx} 
\usepackage{subfigure} 

\usepackage{bbm,amsfonts}
\usepackage{psfrag}
\usepackage{amsbsy}
\usepackage[intlimits]{amsmath} 

\usepackage{epstopdf}

\usepackage[text={6.8in,10in}]{geometry}

\providecommand{\e}[1]{\ensuremath{\cdot 10^{#1}}}

\begin{document}

\begin{titlepage}

\parindent=0pt

\begin{center}
{\huge \bf \sffamily Outbound SPIT Filter with Optimal Performance Guarantees}

\vspace{2cm}
\end{center}

\vspace{3cm}
{\bfseries Tobias Jung}$^1$\\
(corresponding author)\\
{\tt tjung@ulg.ac.be\\}

{\bfseries Sylvain Martin}$^1$\\
{\tt sylvain.martin@ulg.ac.be\\}

{\bfseries Mohamed Nassar$^2$\\}
{\tt mohamed.nassar@inria.fr\\}

{\bfseries Damien Ernst}$^1$\\
{\tt dernst@ulg.ac.be\\}

{\bfseries Guy Leduc}$^1$\\
{\tt guy.leduc@ulg.ac.be\\}

\medskip

$^1$Montefiore Institute (Department of EECS)\\
University of Li\`ege\\
Belgium

\bigskip

$^2$ INRIA Grand Est - LORIA Research Center\\
France
\end{titlepage}

\begin{center}
{\large \bf \sffamily Outbound SPIT Filters with Optimal Performance Guarantees}\\ \vspace{1.5cm}

\medskip
Draft November 30, 2011
\end{center}

\vspace{3cm}

\begin{abstract}
This paper presents a formal framework for identifying and filtering SPIT calls (SPam in Internet
Telephony) in an outbound scenario with provable optimal performance. In so doing, our work is largely
different from related previous work: our goal is to rigorously formalize the problem in terms of mathematical 
decision theory, find the optimal solution to the problem, and derive concrete bounds for its expected loss 
(number of mistakes the SPIT filter will make in the worst case). 
This goal is achieved by considering an abstracted scenario amenable to theoretical analysis, namely  
SPIT detection in an outbound scenario with pure sources. Our methodology is to first define the cost 
of making an error (false positive and false negative), apply Wald's sequential probability ratio test
to the individual sources, and then determine analytically error probabilities such that the resulting 
expected loss is minimized. 
The benefits of our approach are: (1) the method is optimal (in a sense defined in the paper);  
(2) the method does not rely on manual tuning and tweaking of parameters but is completely self-contained 
and mathematically justified; (3) the method is computationally simple and scalable. These are desirable
features that would make our method a component of choice in larger, autonomic frameworks.
\end{abstract}
\vspace*{1.5cm}


\vspace*{0.5cm}
{\bf Short title:} Outbound SPIT Filters with Optimal Performance Guarantees

\newpage

\section{Introduction}
\label{seq:introduction}
Over the last years, Voice over IP (VoIP) has gained momentum as the natural complementary to emails,  
although its adoption is still young. The technologies employed in VoIP are widely similar to those 
used for e-mails and a large portion is actually identical. As a result, one can easily produce hundreds
concurrent calls per second from a single machine, replaying a pre-encoded message as soon as the callee
accepts the call. This application of SPAM over Internet Telephony -- also known as SPIT -- is considered 
by many experts of VoIP as a severe potential threat to the usability of the technology \cite{spider-final-report08}. 
More concerning, many of the defensive measures that are effective against email SPAM do not directly 
translate in SPIT mitigation: unlike with SPAM in emails, where the content of a message is text and 
is available to be analyzed {\em before} the decision is made of whether to deliver it or flag it as SPAM, 
the content of a phone call is a voice stream and is only available when the call is actually answered.

The simplest guard against SPIT would be to enforce strongly authenticated identities 
(maintaining caller identities on a secure and central server) together with 
personalized white lists (allowing only friends to call) and a consent framework  
(having unknown users first ask for permission to get added to the list).
However, this is not supported by the current communication protocols and also 
seems to be infeasible in practice.
Thus a number of different approaches have been previously suggested to address SPIT prevention, 
which mostly derive from experience in e-mail or web SPAM defense. They range from reputation-based 
\cite{Kolan07} and call-frequency based \cite{Shin06} dynamic black-listing, fingerprinting \cite{C06}, 
to challenging suspicious calls by captchas \cite{schlegel06,dtmf-captcha-for-SPIT,quitt:icc}, or 
the use of more sophisticated machine learning. For example, \cite{nassar-SVM,nassar10} suggests SVM for anomaly
detection in call histories, and \cite{wu09} proposes semi-supervised learning, a variant of 
k-means clustering with features optimized from partially labeled data, to cluster and 
discriminate SPIT from regular calls.

These methods provide interesting building blocks, but, in our opinion, suffer from two main
shortcomings. First, they do not provide performance guarantees in the sense that it is difficult 
to get an estimate of the number of SPIT calls that will go through and the number of regular
calls they will erroneously stop. Second, they require a lot of hand-tuning for working well, which
cannot be sustained in today's networks.
 
The initial motivation for this paper was to explore whether there would be ways to design SPIT filters that
would not suffer from these two shortcomings. For this, we start by considering an abstracted scenario amenable to 
theoretical analysis where we make essentially two simplifying assumptions: 
\begin{enumerate}
\item we are dealing with {\em pure source} SPIT detection in an {\em outbound} scenario, 
\item we can extract features from calls (such as, for example, call duration) whose 
distribution for SPIT and regular calls is {\em known in advance}. 
\end{enumerate}
Here, ``outbound scenario'' means that our SPIT detector will be located in, or at the egde of, 
the network where the source resides, and will check all outgoing calls originating from 
within the network. Technically, this means that we are able to easily map calls to sources
and that we can observe multiple calls from each source.
By ``pure source'' we mean that a source either produces only SPIT or only NON-SPIT calls for a
certain observation horizon. 
Under these assumptions, we have been able to design a SPIT filter which requires 
no tuning and no user feedback and which is optimal in a sense that will be defined later in this
paper.

This paper reports on this SPIT filter and is organized in two parts: one theoretical in Section~3 and one 
practical in Section~4. 
The theoretical part starts with Section~\ref{seq:problem statement} describing precisely and in  
mathematical terms the context in which we will design the SPIT filter. Section~\ref{seq:algorithm} 
shows how it is then possible to derive from a simple statistical test a SPIT filter with the 
desired properties and Section~\ref{seq:expectedloss} provides analytical expressions to compute
its performance. Monte Carlo simulations in Section~\ref{seq:example_exponential} and \ref{seq:experiment1}
then examine the theoretical performance of the SPIT filter. 
The practical part starts with Section 4.1 describing how the SPIT filter could be integrated as 
one module into a larger hierarchical SPIT prevention framework. The primary purpose of this section is 
to demonstrate that the assumptions we make in Section 3 are well justified and can be 
easily dealt with in a real world application. (Note that a detailed description of the system architecture
is not the goal of this paper.) For example, Section 4.2 describes how the assumption  
that the distributions for SPIT and NON-SPIT must be known in 
advance can be dealt with using maximum likelihood estimation from labeled prior 
data (which in addition allows us to elegantly address the problem of nonstationary attackers). 
Then, using learned distributions, Section 4.3 demonstrates for data extracted from a large 
database of real-world voice call data that the performance of our SPIT filter in the real world is 
also very good and is in accordance with the performance bounds derived analytically in Section 3.

\section{Related Work}
\label{seq:related work}
To systematically place our work in the context of related prior work, we will have to consider 
it along two axes. The first axis deals with (low-level) {\em detection} algorithms: here we
have to deal with the question on what abstract object we want to work with (e.g., SIP headers, 
stream data, call histories), how to represent this object such that computational algorithms 
can be applied (e.g., what features), and what precise algorithm is applied to arrive at 
a decision, which can be a binary classification (NON-SPIT/SPIT), a score (interpreted
as how likely it is to be NON-SPIT/SPIT), or something else.
The second axis deals with larger SPIT detection {\em frameworks} in which the 
(low-level) detection algorithm is only a small piece. The framework manages and controls the 
complete flow and encompasses detection, countermeasures, and self-healing. 
The formal framework for SPIT filtering we propose in this paper clearly belongs to the first
category and only addresses low-level detection.

The Progressive Multi Gray-leveling (PMGL) proposed in \cite{Shin06} is a low-level detector 
that monitors call patterns and assigns a gray level to each caller based on the number of calls 
placed on a short and long term. If the gray level is higher than a certain threshold, the call is deemed
SPIT and the caller blocked. The PMGL is similar to what we are doing in that it attempts to 
identify sources that are compromised by bot nets in an outbound scenario. The major weakness of 
the PMGL is: (1) that it relies on a weak feature, as spitters can easily adapt and reduce their calling rate to remain
below the detection threshold, and high calling rates can also have other valid causes such as a
natural disaster; (2) that it relies on ``carefully'' hand-tuned detection thresholds to work, 
which makes good performance in the real world questionable and -- in our opinion -- is not a very 
desirable property as it does not come with any worst-case bounds or performance guarantees.  
Our approach is exactly the opposite as it starts from a mathematically justified scenario and
explicitly provides performance guarantees and worst-case bounds. Our approach is also more generic
because it can work with {\em any} feature representation: while we suggest call duration is a better 
choice than call rate, our framework will work with whatever feature representation a network operator
might think is a good choice (given their data). 

In \cite{wu09} a low-level detector based on semi-supervised clustering is proposed. They use a large 
number of call features, and because most of the features become available only after a call is accepted, 
is also primarily meant to classify pure sources (as we do). The algorithm they propose is more 
complex and computationally more demanding than what we propose. In addition, their algorithm 
also relies on hand-tuned parameters and is hard to study analytically; thus again it is impossible to have 
performance guarantees and derive worst-case bounds for it.
Performance-wise it is hard to precisely compare the results due to a different experimental setup, 
but our algorithm compares favorably and achieves a very high accuracy.  

The authors of \cite{nassar-SVM,nassar10} propose to use support vector machines for identifying a varied 
set of VoIP misbehaviors, including SPIT. Their approach works on a different representation of the problem
(call histories) with a different goal in mind and thus is not directly comparable with what we do. While it 
also cannot offer performance guarantees and worst-case bounds, in some respect it is more general 
than what we propose; it also describes both detection and remediation mechanisms in a larger framework. 
  
In SEAL \cite{schlegel06}, the authors propose a complete framework for SPIT prevention which is organized in 
two stages (passive and active). The passive stage performs low-level detection and consists of simple, unintrusive 
and computationally cheap tests, which, however, will only be successful in some cases and can be easily 
fooled otherwise. The purpose of the passive stage is to screen incoming calls and flag those that could
be SPIT. The active stage then performs the more complex, intrusive and computationally expensive tests, which
with very high probability can identify SPIT (these tests more actively interact with the caller). SEAL is very 
similar to what we sketch in Section~4 of this paper. On the other hand, the low-level detection performed in 
SEAL is rather basic: while it is more widely applicable than what we do, it essentially consists only in 
comparing a weighted sum of features against a threshold. As with all the other related work, weights and 
thresholds again need to be carefully determined by hand; and again, since the problem is not modeled 
mathematically, performance guarantees and worst-case bounds are impossible to derive.

Finally, it should be noted that, while the problem of SPIT detection can, in some sense, be related to the problem of anomaly 
detection and prevention of DoS attacks in VoIP networks, for example see the work in \cite{Self-learning08,A07,B08,AA10}, 
it is not the same. The reason why this is not the same is that these security threats are typically specific attacks aimed at disrupting the normal 
operation of the network. SPIT on the other hand operates on the social level and may consist of unwanted advertisements or phishing
attempts -- but not {\em per se} malicious code. As a consequence, techniques from anomaly detection and prevention of DoS attacks cannot be
directly applied to SPIT detection.

\section{A SPIT Filter with Theoretically Optimal Performance}
\label{seq:theory}
This section describes a formal framework for an outbound SPIT filter for which it is possible to prove optimality and provide
performance guarantees. Note that this section is stated from a theoretical point of view. In Section~\ref{seq:real world} 
we outline how one could implement it in a real world scenario.  

\subsection{Problem Statement}
\label{seq:problem statement}

\begin{figure}
\begin{center}
\includegraphics[width=8cm]{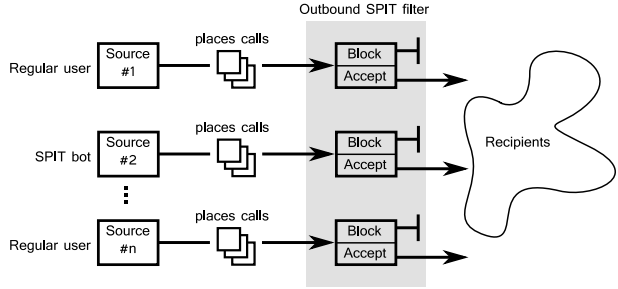}
\end{center}
\caption{Sketch of the simplified problem. The SPIT filter operates in an outbound fashion and checks all outgoing calls originating 
from within the network. The filter treats the source of a call as being pure over a certain observation horizon. In this scenario, 
each source (which will correspond to a registered user) will try to sequentially place calls over time which are either all SPIT 
(if the source is compromised by a SPIT bot) or all NON-SPIT (if the source is a regular user). The SPIT filter then has to decide
for each source individually if, given the calls that originated from that source in the past, it is a regular user or a SPIT bot.}
\label{fig:1}
\end{figure}

As shown in Figure~\ref{fig:1}, we assume the following situation: the SPIT filter receives and monitors 
incoming calls from a number of different sources 
(in real-world operation the SPIT filter will be placed in the outbound network, but implementation details such as this will 
be ignored in this theoretical section).
The sources are independent 
from each other and will each place numerous calls over time. We assume that, over a given 
observation horizon, each source will either {\em only produce regular calls} or will 
{\em only produce SPIT calls}. As the sources are independent, we can run a separate instance 
of the SPIT filter for each and thus in the following will only deal with the case of a single 
source.

Every time a call arrives at the filter, the filter can do one of two things: (i) accept the call and pass it
on the recipient or (ii) block the call. Each call will be associated with certain features and we assume
that only if a call is accepted, we can observe the corresponding feature (e.g., call duration). The fundamental
assumption we then make is that the distribution over these features will be different depending on whether a
call is a SPIT call or a regular call, and that these distributions are known in advance. 

Our goal is to decide, after observing a few calls, whether or not the source sends out SPIT. More precisely, 
we look for a decision policy that initially accepts all calls, thus refining the belief about whether or 
not the source is SPIT, and then at some point decides to either block or accept all future calls from 
the source. Seen as a single-state decision problem over time, the SPIT filter has 
three possible actions: (1) accept the next call, which reveals its features and thus refines the belief about 
the type of the source, (2) block all future calls, and (3) accept all future calls.
The last two actions immediately stop the decision-making process and, if the decision was 
wrong (that is, deciding to accept when in truth the source is SPIT, or deciding to block
when in truth the source is NON-SPIT), will incur a terminal loss proportional to the number 
of remaining calls. In addition, we also have a per step cost during the initial exploration
if the source is of type SPIT (for erroneously accepting SPIT calls).

In doing so, we arrive at a well defined concept of loss. 
Within the framework outlined above, every conceivable SPIT filter algorithm will have a 
performance number: its expected loss.
The particular SPIT filter that we are going to describe below will be one that minimizes
this expected loss.  

Note that the expected loss is an example for the typical exploration vs exploitation 
dilemma.\footnote{Note that while at first glance the scenario described might bear a strong 
resemblance to traditional 2-armed bandit formulations, the major difference is that here we 
assume that the precise form and parameters of the underlying distributions are known in 
advance whereas in typical bandit scenarios this is not the case \cite{Robbins52}.} On the 
one side, since the decision to accept or block all future calls is 
terminal, we want to be very certain about its correctness to avoid making an expensive error. 
On the other side, as long as we are observing we are automatically accepting all calls and 
thus will increase our loss if the source turns out to produce SPIT calls. To minimize our expected 
loss, we therefore also want to observe as few samples as possible.

To address the problem mathematically, we employ Wald's sequential probability ratio test for simple 
hypotheses introduced in \cite{Wald43}. The sequential probability ratio test (SPRT) has the 
remarkable property that among all sequential tests procedures it minimizes the expected number
of samples for a given level of certainty and regardless of which hypothesis is true (the optimality 
of SPRT was proved in \cite{WaldWolf48}). In addition, the SPRT comes with bounds for the expected
stopping time and thus allows us to derive concrete expressions for the expected loss as a function
of the characteristics of the particular problem (meaning we can express the loss as a function 
of the parameters of the distribution for SPIT or NON-SPIT). Finally, SPRT requires only 
simple algebraic operations to carry out and thus is easy to implement and computationally
cheap to run.

\newcommand{\pspit}{p_{\text{SPIT}}}
\newcommand{\pnonspit}{p_{\text{NON-SPIT}}}

\subsection{SPIT Detection via the SPRT }
\label{seq:algorithm}
The SPRT is a test of one simple statistical hypothesis against another which operates in an
online fashion and processes the data sequentially. At every time step a new observation is 
processed and one of the following three decisions is made: (1) the hypothesis being tested 
is accepted, (2) the hypothesis being tested is rejected and the alternative hypothesis is
accepted, (3) the test statistic is not yet conclusive and further observations are necessary.
If the first or the second decision is made, the test procedure is terminated. Otherwise the process
continues until at some later time the first or second decision is made.
 
Two kind of misclassification errors may arise: decide to accept calls when source is SPIT, or
decide to block all future calls when source is NON-SPIT. 
Different costs may be assigned to each kind, 
upon which the performance optimization process described in Section~\ref{seq:expectedloss} is built. 

To model the SPIT detection problem with the SPRT, we now proceed as follows: Assume we can make
sequential observations from one source of {\em a priori} unknown type SPIT or NON-SPIT. Let $x_t$
denote the features of the $t$-th call we observe, modeled by random variable $X_t$. The $X_t$ are 
i.i.d. with common distribution (or density) $p_X$. The calls all originate from one source
which can either be of type SPIT with distribution $\pspit(x)=p(x|\text{source=SPIT})$ or of type 
NON-SPIT with distribution $\pnonspit(x)=p(x|\text{source=NON-SPIT})$. Initially, the type of the source
we are receiving calls from is not known; in absence of other information we have to assume that both types 
are equally likely, thus the prior would be $p(\text{SPIT})=p(\text{NON-SPIT})=\frac{1}{2}$.
In order to learn the type of the source, we observe calls $x_1,x_2,\ldots$ and test the hypothesis
\begin{equation}
H_0: \, p_X=\pspit \qquad \text{versus} \qquad H_1: \, p_X=\pnonspit. 
\end{equation}
(Note again that in this formulation we assume that the densities $\pspit$ and $\pnonspit$ are both known so that
we can readily evaluate the expression $p(x|\text{source=SPIT})$ and $p(x|\text{source=NON-SPIT})$
for any given $x$.)

At time $t$ we observe $x_t$. Let
\begin{equation}
\lambda_t:=\frac{p(x_1,\ldots,x_t|\text{NON-SPIT})}{p(x_1,\ldots,x_t|\text{SPIT})}=
\prod_{i=1}^t \frac{p(x_i|\text{NON-SPIT})}{p(x_i|\text{SPIT})}
\end{equation}
be the ratio of the likelihoods of each hypothesis after $t$ observations $x_1,\ldots, x_t$. 
Since the $X_i$ are independent we can factor the joint distribution on the left side to obtain 
the right side. In practice it will be more convenient for numerical reasons to work with the
log-likelihoods. Doing this allows us to write a particular simple recursive update for the log-likelihood
ratio $\Lambda_t:=\log \lambda_t$, that is
\begin{equation}
\Lambda_t:=\Lambda_{t-1} + \log \frac{p(x_t|\text{NON-SPIT})}{p(x_t|\text{SPIT})}.
\label{eq:loglike}
\end{equation}
After each update we examine which of the following three cases applies and act accordingly:
\begin{align}
A<\lambda_t<B & \Longrightarrow \text{ continue monitoring}  \label{eq:sprtdecision}\\
\lambda_t\ge B & \Longrightarrow \text{ accept } H_1 \text{ (decide NON-SPIT)} \\
\lambda_t\le A & \Longrightarrow \text{ accept } H_0 \text{ (decide SPIT)}
\end{align}
Thresholds $A$ and $B$ with $0<A<1<B<\infty$ depend on the desired accuracy or error probabilities 
of the test:
\begin{align}
\alpha:= & P\{\text{accept } H_1 \,|\, H_0 \text{ true} \}
          =  P\{ \text{decide NON-SPIT} \,|\, \text{source=SPIT}\} \\
\beta:= & P\{\text{reject } H_1 \,|\, H_1 \text{ true} \} 
         = P\{ \text{decide SPIT} \,|\, \text{source=NON-SPIT}\}. 
\end{align}
Note that $\alpha$ and $\beta$ need to be specified in advance such that certain accuracy requirements 
are met (see next section where we consider the expected loss of the procedure). The threshold values $A$
and $B$ and error probabilities $\alpha$ and $\beta$ are connected in the following way
\begin{equation}
\beta \le A(1-\alpha) \qquad \text{and} \qquad \alpha\le(1-\beta)/ B.
\end{equation}
Note that the inequalities arise because of the discrete nature of making observations (i.e., at $t=1,2,\ldots$) 
which results in $\lambda_t$ not being able to hit the boundaries exactly. In practice we will neglect this
excess and treat the inequalities as equalities:
\begin{equation}
A=\beta / (1-\alpha) \qquad \text{and} \qquad B=(1-\beta)/ \alpha.
\end{equation}

Let $T$ be the random time at which the sequence of the $\lambda_t$ leaves the open interval $(A,B)$ and a 
decision is made that terminates the sampling process. (Note that stopping time $T$ is a random quantity due 
to the randomness of the sample generation.) The SPRT provides the following pair of inequalities (equalities)
for the expected stopping time (corresponding to each of the two possibilities)
\begin{align}
E_{X_i \sim \pspit}[T] & = \frac{1}{\kappa_0} \left( \alpha \log \frac{1-\beta}{\alpha} 
+ (1-\alpha) \log \frac{\beta}{1-\alpha} \right) \label{eq:stopping_spit}\\
E_{X_i \sim \pnonspit}[T] & = \frac{1}{\kappa_1} \left( \beta \log \frac{\beta}{1-\alpha} 
+ (1-\beta) \log \frac{1-\beta}{\alpha} \right) \label{eq:stopping_nonspit}.
\end{align}
The constants $\kappa_i$ with $\kappa_0<0<\kappa_1$ are the Kullback-Leibler information numbers defined 
by
\begin{align}
\kappa_0 &= E_{x\sim \pspit} \left[ \log \frac{p(x|\text{NON-SPIT})}{p(x|\text{SPIT})} \right] \label{eq:k0} \\
\kappa_1 &= E_{x\sim \pnonspit} \left[ \log \frac{p(x|\text{NON-SPIT})}{p(x|\text{SPIT})} \right]\label{eq:k1}. 
\end{align}
The constants $\kappa_i$ can be interpreted as a measure of how difficult it is to distinguish between
$\pspit$ and $\pnonspit$. The smaller they are the more difficult is the problem.

\subsection{Theoretical Performance of the SPIT Filter}
\label{seq:expectedloss}
We will now look at the performance of our SPIT filter and derive expressions for its expected loss. 
Let us assume we are going to receive a finite number $N$ of calls and that $N$ is sufficiently large
such that the test will always stop before the calls are exhausted (the case where we observe fewer
samples than what the test demands will be ignored for now, e.g., see the truncated SPRT \cite{Wald43}).

How does the filter work? At the beginning all calls are accepted (observing samples $x_i$) until
the test becomes sufficiently certain about its prediction. Once the test becomes sufficiently
certain, based on the outcome the filter implements the following simple policy: if the test returns
that the source is SPIT then all future calls from it will be blocked. If the test says that the
source is NON-SPIT then all future calls from it will be accepted. Since the decision could be
wrong, we define the following cost matrix (per call):

\begin{center}
{\small
\begin{tabular}{|l|c|c|}
\hline
& Source=SPIT & Source=NON-SPIT \\
\hline
Accept call & $c_0$ & $0$ \\
Block call & $0$ & $c_1$ \\
\hline
\end{tabular}
}
\end{center} 

Let $L$ denote the loss incurred by this policy (note that $L$ is a random quantity). To compute the expected loss, we 
have to divide $N$ into two parts: the first part from $1$ to $T$ 
corresponds to the running time of the test where all calls are automatically accepted ($T<N$
being the random stopping time with expectation given in Eqs.~\eqref{eq:stopping_spit}-\eqref{eq:stopping_nonspit}), 
the second part from $T+1$ to $N$ corresponds to the time after the test.

If $H_0$ is true, that is, the source is SPIT, the loss $L$ will be the random quantity
\begin{equation}
L|\text{source=SPIT} \ = \ c_0 T + \alpha c_0 (N-T)
\end{equation}  
where $c_0 T$ is the cost of the test, $\alpha$ the probability of making the wrong decision,
and $c_0 (N-T)$ the cost of making the wrong decision for the remaining calls.
Taking expectations gives
\begin{equation}
E_{X_i\sim\pspit}[L] = \alpha c_0 N + c_0 (1-\alpha) E_{X_i\sim\pspit}[T].
\end{equation}

Likewise, if $H_1$ is true, that is, the source is NON-SPIT, our loss will be the random quantity
\begin{equation}
L|\text{source=NON-SPIT} \ = \ 0 \cdot T + \beta c_1 (N-T)
\end{equation} 
where $0$ is the cost of the test (because accepting NON-SPIT is the right thing to do), $\beta$ 
the probability of making the wrong decision, and $c_1 (N-T)$ the cost of making the wrong
decision for the remaining calls. Taking expectation gives
\begin{equation}
E_{X_i\sim\pnonspit}[L] = \beta c_1 (N- E_{X_i\sim\pnonspit}[T]).
\end{equation}

The total expected loss takes into consideration both cases and is simply
\begin{equation}
E[L]=p(\text{SPIT}) \cdot E_{X_i\sim\pspit}[L] + p(\text{NON-SPIT}) \cdot E_{X_i\sim\pnonspit}[L].
\end{equation}
For the case that both priors are equal, we have 
\begin{multline}
E[L]=\frac{1}{2} \Bigl\{ N(\alpha c_0 + \beta c_1) + \log \frac{1-\beta}{\alpha}
\Bigl( \frac{c_0 \alpha (1-\alpha)}{\kappa_0} - \frac{c_1 \beta (1-\beta)}{\kappa_1} \Bigr) \\
+
\log \frac{\beta}{1-\alpha} 
\Bigl( \frac{c_0 (1-\alpha)^2}{\kappa_0} - \frac{c_1 \beta^2}{\kappa_1} \Bigl)
\Bigr\}
\label{eq:totalloss}
\end{multline}

Now let us consider the situation where we want to apply the filter in practice. For any given
problem distributions $\pspit$, $\pnonspit$, the number of calls $N$ and the cost of mistakes $c_0, c_1$
are specified in advance. To be able to run the test, the only thing left is how to set the 
remaining parameters $\alpha,\beta$. However, looking at Eq.~\eqref{eq:totalloss} we see that, given
all the other information, the expected loss will be a function of $\alpha,\beta$. Thus one way
of choosing $\alpha,\beta$ would be to look for that setting $\alpha^*,\beta^*$ that will minimize
the expected loss under the given problem specifications (i.e., distributions $\pspit,\pnonspit$ and cost $c_0,c_1$ ). 
Of course, Eq.~\eqref{eq:totalloss} is nonlinear in $\alpha$ and $\beta$ and cannot be minimized analytically 
and in closed form. Instead, we will have to employ iterative techniques to approximately determine $\alpha^*, \beta^*$ or solve a 
simplified problem (e.g., use linearization).

\subsection{Example: Exponential Duration Distribution}
\label{seq:example_exponential}

For the following numerical example we assume that $\pspit$ and $\pnonspit$ are both
exponential distributions with parameters $\lambda_0,\lambda_1 >0$, that is, 
are given by 
\begin{equation}\label{eq:expdist}\begin{split}
p(x|\text{SPIT}) & =\lambda_0 \exp(-\lambda_0x)\\
p(x|\text{NONSPIT}) & =\lambda_1 \exp(-\lambda_1 x)
\end{split}
\end{equation}
for $x>0$.
While this example is primarily meant to illustrate the behavior of a SPRT-based SPIT filter 
theoretically, it is not an altogether unreasonable scenario to assume for a real world SPIT
filter. For example, we could assume that the only observable feature $x$ of (accepted) calls is 
their duration (see Section~\ref{seq:real world}).
In this case SPIT calls will have a shorter duration than regular calls because
after a callee answers the call, they will hang up as soon as they realize it is SPIT. The majority
of regular calls on the other hand will tend to have a longer duration. While this certainly 
simplifies the situation from the real world (e.g., it is generally assumed that call duration follows
a more complex and heavy-tailed distribution \cite{Duffy94statisticalanalysis}), we can imagine that both durations can be
modeled by an exponential distribution with an average (expected) length of SPIT calls
of $1/\lambda_0$ minutes and an average length of NON-SPIT calls of $1/\lambda_1$ minutes 
($\frac{1}{\lambda_1}>\frac{1}{\lambda_0}$).

First, let us consider the expected stopping time from Eqs.~\eqref{eq:stopping_spit}-\eqref{eq:stopping_nonspit}. 
From Eqs.~\eqref{eq:k0}-\eqref{eq:k1} we have that the 
Kullback-Leibler information number for Eq.~\eqref{eq:expdist} is given by
\begin{align}
\kappa_0 &= \int_0^{\infty} \log \left[
\frac{\lambda_1 \exp(-\lambda_1 x)}{\lambda_0 \exp(-\lambda_0 x)}\right] 
\cdot \lambda_0 \exp(-\lambda_0 x) dx \nonumber \\
&=\log \frac{\lambda_1}{\lambda_0} \int_0^{\infty} \lambda_0 \exp(-\lambda_0 x) dx +
+(\lambda_0 - \lambda_1) \int_0^{\infty} x \cdot \lambda_0 \exp(-\lambda_0 x) dx \nonumber \\
&=\log \frac{\lambda_1}{\lambda_0} + 1 - \frac{\lambda_1}{\lambda_0}
\end{align}
(On the second line, the first integral is an integral over a density and thus is equal to one; the 
second integral is the expectation of $\pspit$ and thus is equal to $1/\lambda_0$.)
Similarly we obtain for $\kappa_1$ the expression
\begin{equation}
\kappa_1=\log \frac{\lambda_1}{\lambda_0} - 1 + \frac{\lambda_0}{\lambda_1}.
\end{equation}
Note that for more complex forms of distributions we may no longer be able to evaluate $\kappa_i$ in 
closed form.

As we can see, $\kappa_i$ only depends on the ratio $\lambda_1/\lambda_0$. Thus for fixed accuracy
parameters $\alpha,\beta$ the expected stopping time in Eqs.~\eqref{eq:stopping_spit}-\eqref{eq:stopping_nonspit}
will also only depend on the ratio $\lambda_1/\lambda_0$. The closer the ratio is to zero, the fewer samples will 
be needed (the problem becomes easier); the closer the ratio is to one, the more samples will be needed (the problem becomes harder). 
Of course this result is intuitively clear: the ratio $\lambda_1/\lambda_0$ determines how similar the distributions are. 

In Table~\ref{table:1} we examine numerically the impact of the difficulty of the problem, in terms of the ratio
$\lambda_1/\lambda_0$, on the expected number of samples until stopping for different settings of
accuracy $\alpha,\beta$. For instance, an average NON-SPIT call duration of 2 minutes as opposed to an average duration of SPIT calls of 12\,s leads to $\lambda_1/\lambda_0=0.1$, and distributions that are sufficiently dissimilar to arrive with high accuracy at the correct decision within a very short observation horizon: with accuracy $\alpha,\beta=0.001$, the filter has to observe on the average 1.0 calls if the source is NON-SPIT and 4.9 calls if the source is SPIT to make the correct decision in at least 99.9\% of all cases. (Notice that the stopping time is not symmetric.) On the other hand, with $\lambda_1/\lambda_0>0.5$ the similarity between SPIT and NON-SPIT becomes too strong, which is an indication that other/more features should be tried.

\begin{table*}[t]
%
\begin{center}
{\small
\begin{tabular}{|c|rr||rr|rr|rr|}
\hline
& & & \multicolumn{2}{|c|}{$\alpha,\beta=0.05$} & \multicolumn{2}{|c|}{$\alpha,\beta=0.01$} & \multicolumn{2}{|c|}{$\alpha,\beta=0.001$} \\
\hline
$\lambda_1/\lambda_0$ & $\kappa_0$ & $\kappa_1$ & $E_{\text{SPIT}}[T]$ & $E_{\text{NON}}[T]$ & $E_{\text{SPIT}}[T]$ & $E_{\text{NON}}[T]$ &$E_{\text{SPIT}}[T]$ & $E_{\text{NON}}[T]$ \\
\hline
0.99 & -0.00005 &	0.00005	& 52646.2	& 52294.7	& 89463.4	& 88865.9	& 136938.9	& 136024.5\\
0.95 & -0.00129	& 0.00133	& 2049.0	& 1980.1	& 3481.9	& 3364.9	& 5329.7	& 5150.5\\
0.90 & -0.00536	& 0.00575	&  494.3	&  460.8	&  840.0	&  783.0	& 1285.8	& 1198.6\\
0.70 & -0.05667	& 0.07189	&   46.7	&  36.8	 & 79.4	   & 62.6	 & 121.6	  & 95.8\\
0.50 & -0.19314	& 0.30685	&   13.7	&   8.6	 & 23.3	   & 14.6	 &  35.6	  & 22.4\\
0.30 & -0.50397	& 1.12936	&    5.2	&   2.3	 &  8.9	   & 3.9   & 	13.6	  &  6.1\\
0.10 & -1.40258	& 6.69741	&    1.8	&   0.3	 &  3.2	   & 0.6	 &   4.9	  &  1.0\\
0.01 & -3.61517	& 94.39486 &   0.7	&  $<$0.1	 &  1.2	   &$<$0.1	 &   1.9	  &  0.1\\
\hline
\end{tabular}
}
\end{center}
\caption{How does the difficulty of the problem, expressed in terms of the ratio
$\lambda_1/\lambda_0$, affect the expected number of samples until stopping 
$E_{\text{SPIT}}[T]$ and $E_{\text{NON-SPIT}}[T]$ for different settings of 
the accuracy parameters $\alpha,\beta$.}
\label{table:1}
\end{table*}

\begin{figure}
\begin{center}
\includegraphics[width=8cm]{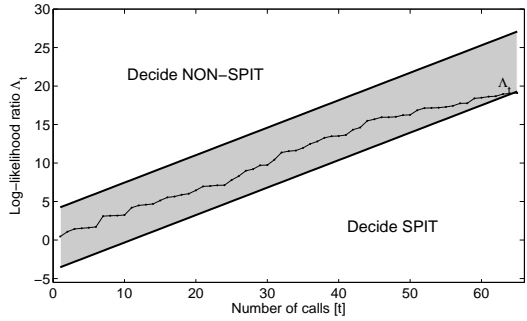}
\end{center}
\caption{An example run of SPRT. The log-likelihood ratio $\Lambda_t$ starts at $t=1$ in the region between 
the decision boundaries and, with each new sample observed from the unknown source, performs a random walk 
over time. Eventually it will cross over one of the decision boundaries and either enter the region marked 
``decide SPIT'' or enter the region marked ``decide NON-SPIT''. 
}
\label{fig:2}
\end{figure}

Next we will compute the log-likelihood ratio $\Lambda_t$. From Eq.~\eqref{eq:loglike} we have
\begin{equation}
\Lambda_t = \sum_{i=1}^t \log \frac{p(x_i|\text{NON-SPIT})}{p(x_i|\text{SPIT})} 
%
%
 = \sum_{i=1}^t \left[ \log \frac{\lambda_1}{\lambda_0} + (\lambda_0 -\lambda_1) x_i \right].
\end{equation}
The decision regions for the SPRT from Eq.~\eqref{eq:sprtdecision} are thus
\begin{equation}
\log \frac{\beta}{1-\alpha} < 
t \cdot \log \frac{\lambda_1}{\lambda_0} + (\lambda_0 -\lambda_1) \sum_{i=1}^t x_i  <
\log \frac{1-\beta}{\alpha}
\end{equation}
or, equivalently,
\begin{equation}
\log \frac{\beta}{1-\alpha} + t \cdot \left(  \log \frac{\lambda_0}{\lambda_1} \right) < 
(\lambda_0 -\lambda_1) \sum_{i=1}^t x_i  <
\log \frac{1-\beta}{\alpha} + t \cdot \left(  \log \frac{\lambda_0}{\lambda_1} \right).
\end{equation}
From the latter we can see that the boundaries of the decision regions are straight and parallel
lines (as a function $t$ of samples). Running the SPRT can now be graphically visualized as shown 
in Figure~\ref{fig:2}: the log-likelihood ratio $\Lambda_T$ starts for $t=1$ in the middle region between the
decision boundaries and, with each new sample it observes from the unknown source, does a random walk over 
time. Eventually it will cross over one of the lines 
after which the corresponding decision is made. For a fixed value of $\alpha,\beta$, changing the ratio
$\lambda_0/\lambda_1$ changes the slope of the decision boundaries. For a fixed value of $\lambda_0, \lambda_1$,
changing the accuracy $\alpha,\beta$ shifts the decision boundaries upward and downward. 

\subsection{Experiment: perfectly known distribution}
\label{seq:experiment1}
\begin{figure}[t!p]
\begin{center}
\includegraphics[width=0.49\textwidth]{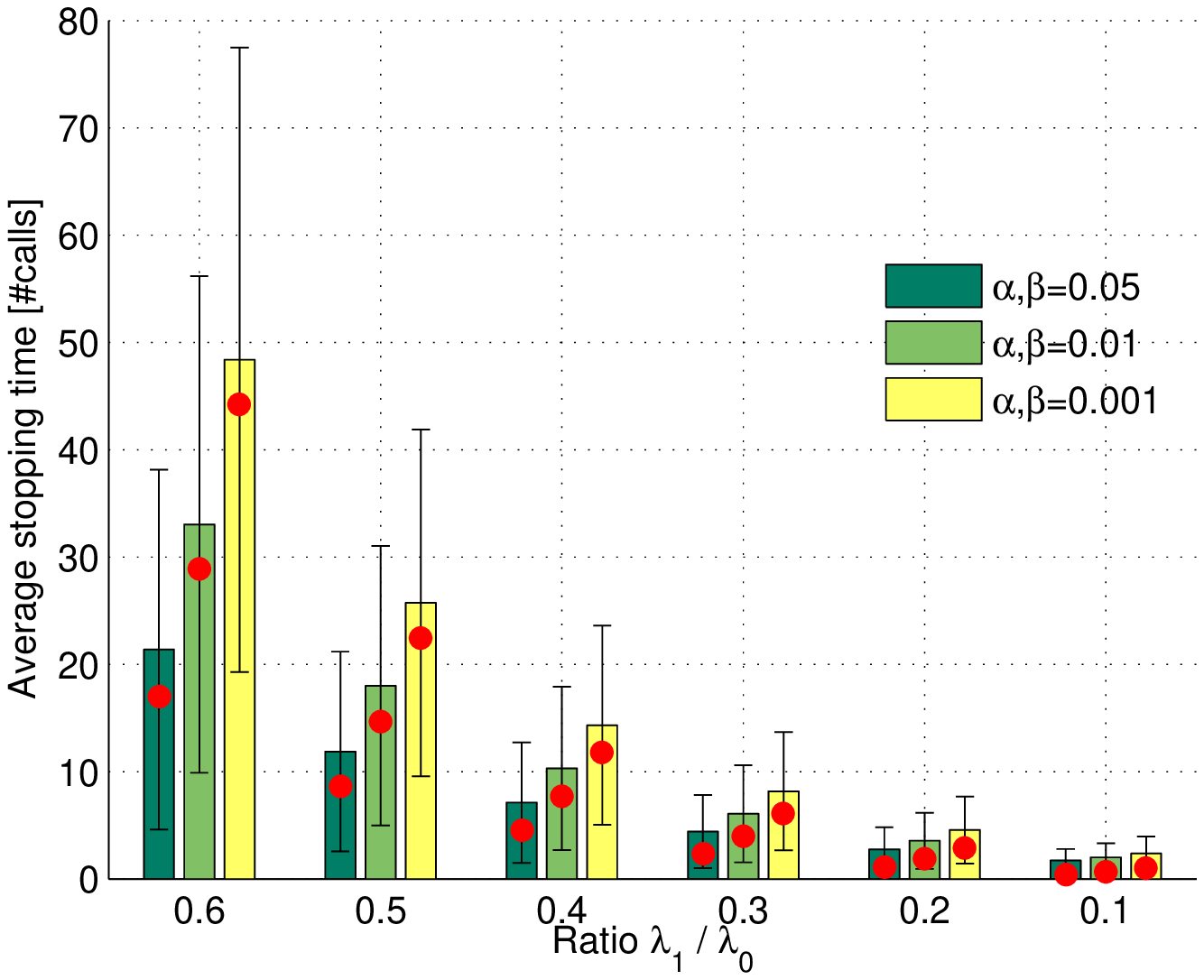}
\includegraphics[width=0.49\textwidth]{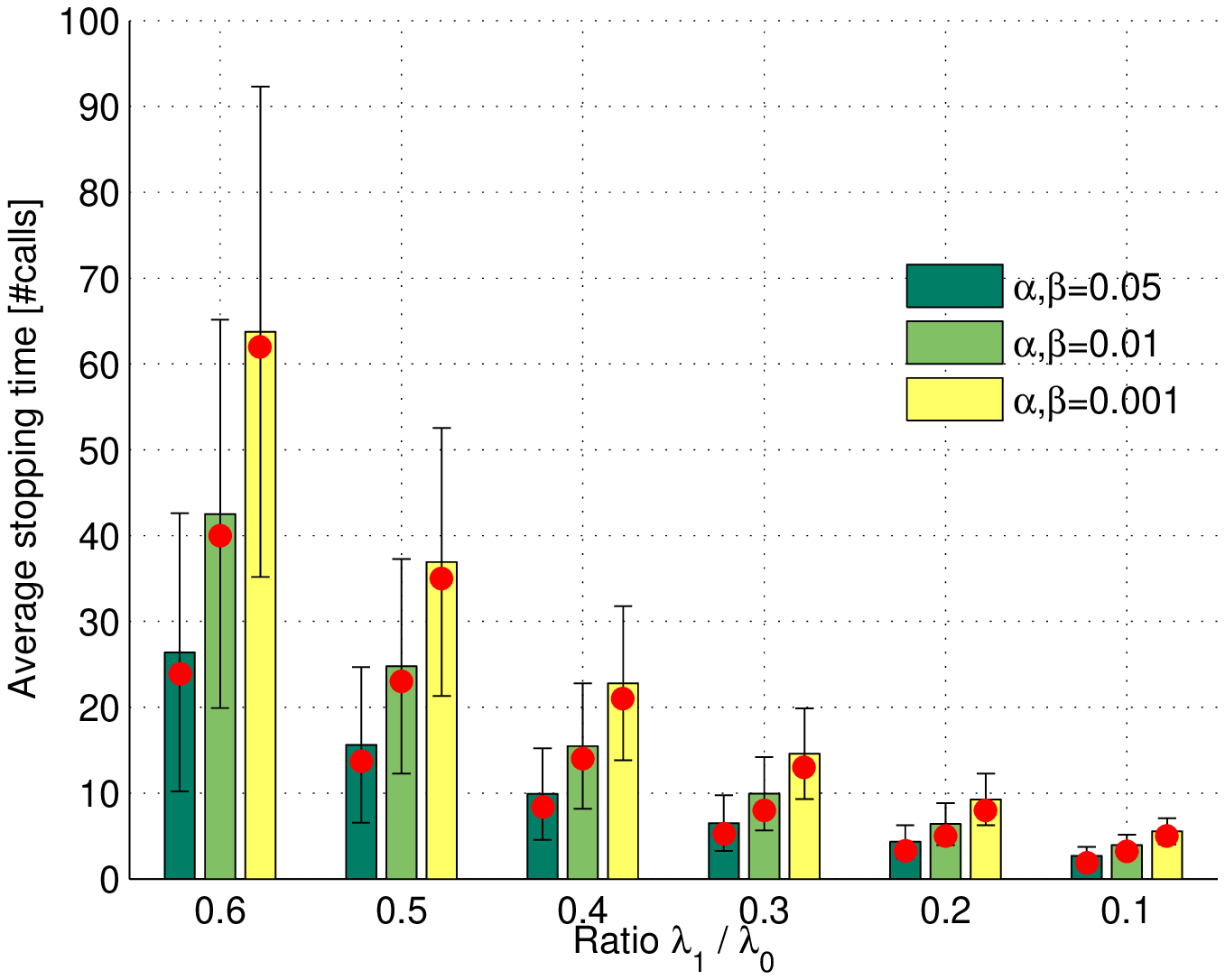}
Average stopping time if source is SPIT (left) and NON-SPIT (right) \\

\includegraphics[width=0.49\textwidth]{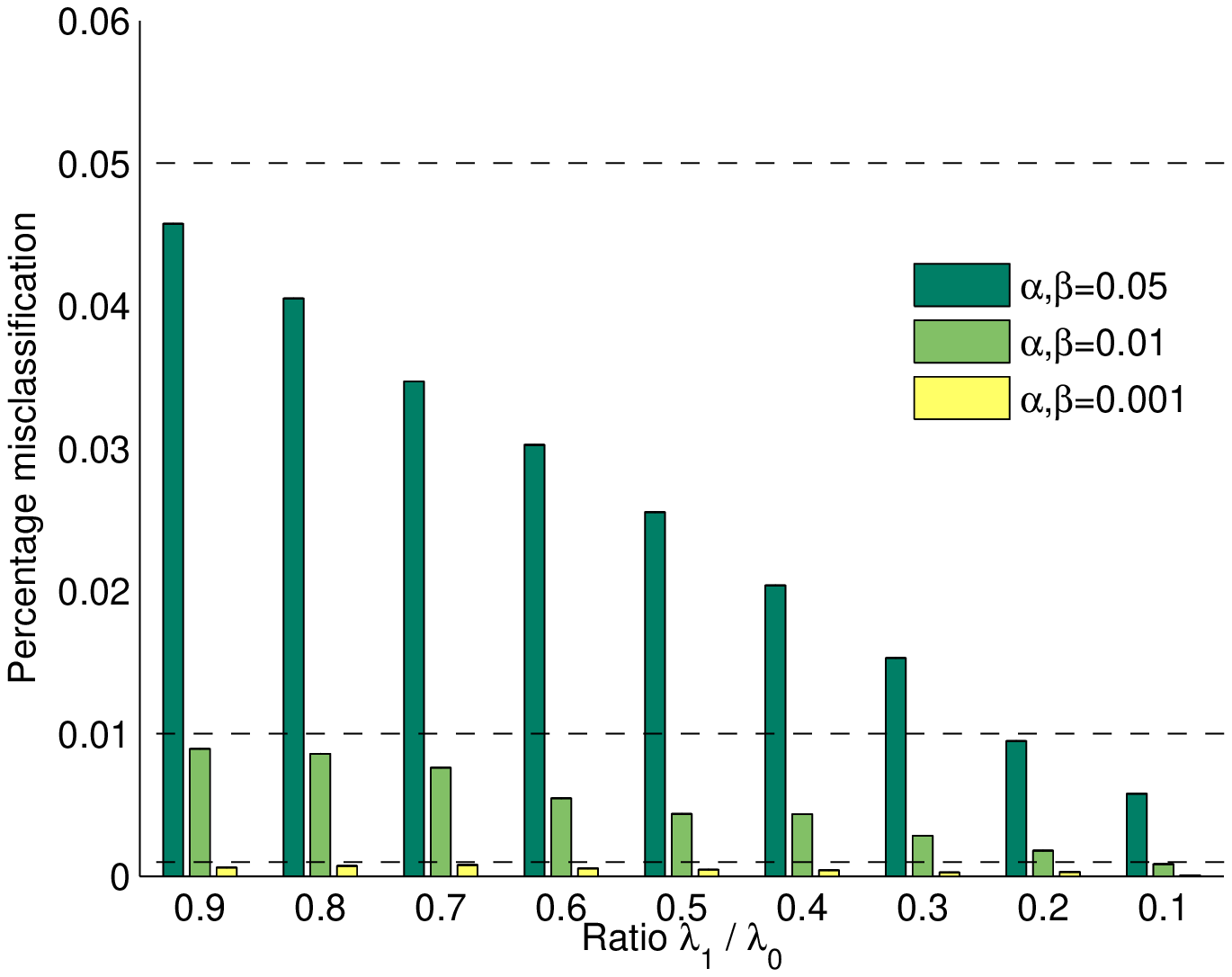} 
\includegraphics[width=0.49\textwidth]{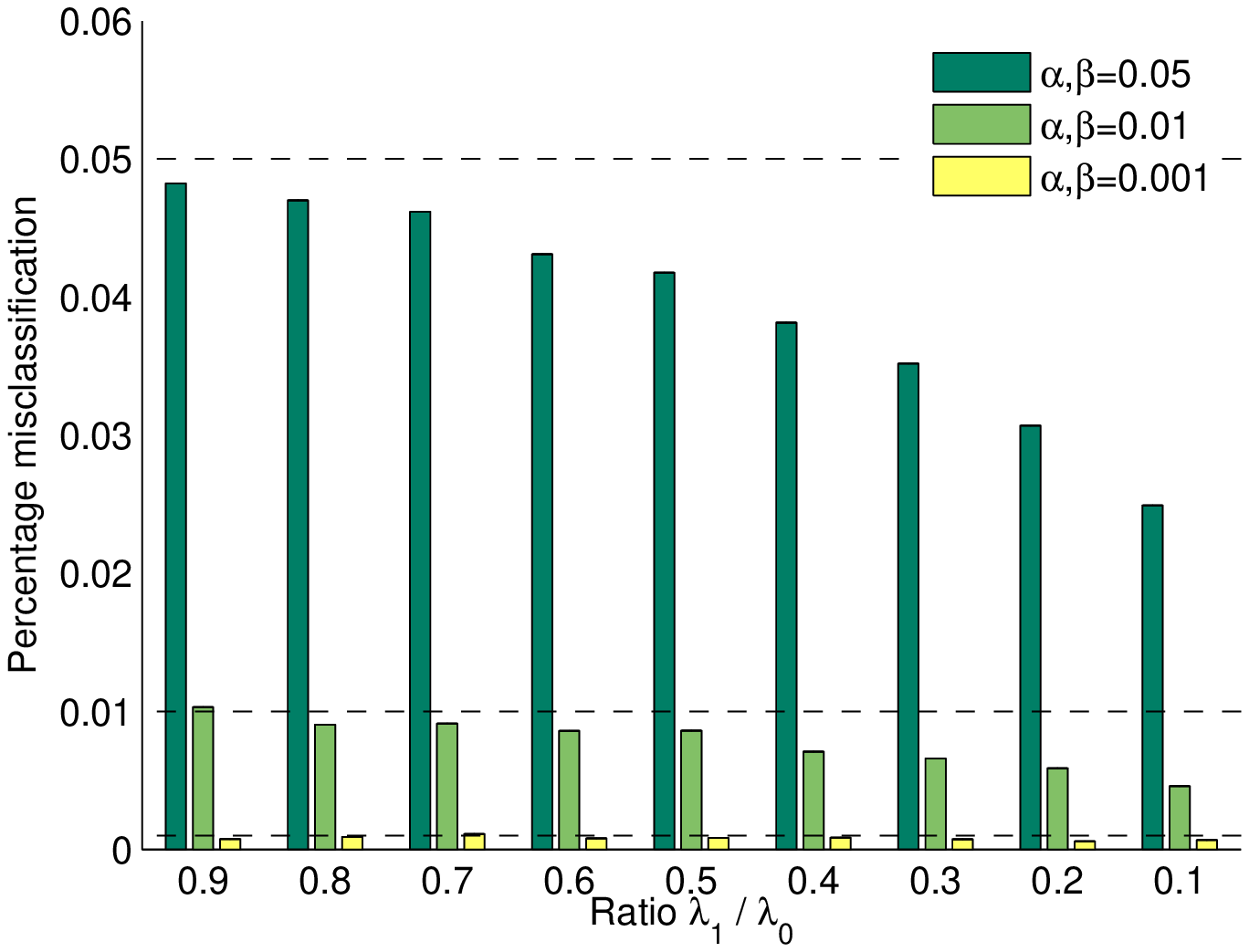}
Error rate if source is SPIT (left) and NON-SPIT (right)
\end{center}
\caption{Monte Carlo simulation of SPRT with the results averaged over 50,000 independent 
runs and error bars denoting one standard deviation.  The top row shows the average number of calls 
necessary before  SPRT stops for various settings of accuracy $\alpha,\beta$. The red dots indicate 
the respective expected stopping time. The bottom row shows the proportion of SPRT
ending up making the wrong decision (that is, accepting SPIT or blocking NON-SPIT) for various settings of 
accuracy $\alpha,\beta$ (shown as horizontal lines). As can be seen, in many instances the height of the bars is 
notably below the corresponding horizontal line, meaning that the actual error rate can be much smaller than what
the accuracy parameters $\alpha,\beta$ would suggest, which are merely an upper bound. }
\label{fig:3}
\end{figure}

To examine the (theoretical) performance of SPRT, we ran a large number of Monte Carlo simulations for various
settings of problem difficulty $\lambda_1/\lambda_0$ ($\lambda_0$ was set to $1$, $\lambda_1$ was varied between
$0.9$ and $0.1$) and accuracy $\alpha,\beta$ ($\alpha,\beta=0.05$, $\alpha,\beta=0.01$, $\alpha,\beta=0.001$). For
each setting we performed 50,000 independent runs and recorded, for each run, how many samples were necessary for SPRT
to reach a decision and whether or not that decision was wrong. The experiment examines separately the case where 
the source is SPIT and NON-SPIT. The result of the simulation is shown in Figure~\ref{fig:3} 
and confirms the expected stopping time computed analytically in Table~\ref{table:1}. In addition, the
results show that the actual number of mistakes made (the height of the bars in the figure) is in many cases notably 
smaller than the corresponding error probabilities $\alpha,\beta$ (the dashed horizontal lines in the figure), which are 
merely upper bounds.

Next, we examine the performance of the SPRT-based SPIT filter with loss function from Eq.~\eqref{eq:totalloss}
in Table~\ref{tab:2}. This time the accuracy parameters $\alpha,\beta$ were not set by hand but chosen to be 
optimal with respect to the loss function Eq.~\eqref{eq:totalloss}  (the minimization was carried out using MATLAB's interior
point solver for constrained problems). To avoid getting overly large stopping times for $\lambda_1/\lambda_0$ close to $1$, $\alpha,\beta>0$
were constrained to be greater than $10^{-5}$ during the minimization. The priors in Eq.~\eqref{eq:totalloss} for $\pspit$ and $\pnonspit$ were both set 
to $1/2$. In the experiment we examine the effect of setting different values for the difficulty $\lambda_1 /\lambda_0$, the cost of blocking 
NON-SPIT $c_1=k\cdot c_0$, and the number of calls ($N=500$ or $N=5,000$). For each such setting, we ran 100,000 independent simulation
trials and computed the average of the result. The table shows that either increasing $c_1$ with respect to $c_0$ or increasing the number of 
calls $N$ will make the optimizer prefer smaller values for $\beta$ to avoid making costly errors (which is proportional to $N$ and $c_1$),
which in turn increases the number of samples necessary for SPRT to stop. The table also shows, as we have already seen 
before, that the numbers of actual misclassifications is below the respective values of the error probabilities 
$\alpha^*$ and $\beta^*$. More importantly, the number itself is very small: out of 100,000 trials the number of times the SPIT filter 
made the wrong decision varied between 0 and 5, which is less than $0.005\%$.

\begin{table*}[t]
\centering
{\scriptsize
\begin{tabular}{|c|c|ccc||cc|}
\hline
& & \multicolumn{5}{|c|}{source=SPIT} \\
\hline
& & \multicolumn{3}{|c||}{N=500} & \multicolumn{2}{|c|}{N=5000} \\
\hline
$\lambda_1/ \lambda_0$ & & $c_1=c_0$ & $c_1=10c_0$ & $c_1=100c_0$ & $c_1=c_0$ & $c_1=10c_0$ \\
\hline
                             & $\beta^*$  & 0.0014 & 0.0001 & 0.0001 & 0.0001 & 0.0001\\
                   0.1     & $nErr$  &      1 &      1 &      2 &      1 &      1\\
                             & $T$        &   5.31 &   6.95 &   7.20 &   6.93 &   7.20\\
\hline 
                             & $\beta^*$  & 0.0024 & 0.0002 & 0.0001 & 0.0002 & 0.0001\\
                   0.2     & $nErr$  &      3 &      2 &      5 &      2 &      1\\
                             & $T$        &   8.14 &  11.01 &  12.12 &  11.01 &  12.11\\
\hline
                             & $\beta^*$  & 0.0040 & 0.0004 & 0.0001 & 0.0004 & 0.0001\\
                   0.3    & $nErr$  &      3 &      3 &      2 &      3 &      3\\
                             & $T$        &  11.82 &  16.37 &  19.17 &  16.41 &  19.11\\
\hline
                             & $\beta^*$  & 0.0065 & 0.0006 & 0.0001 & 0.0006 & 0.0001\\
                   0.4    & $nErr$  &      4 &      5 &      0 &      5 &      5\\
                             & $T$        &   5.31 &   6.95 &   7.20 &   6.93 &   7.20\\
\hline
\end{tabular}
}
\caption{Empirical performance of the SPRT-based SPIT filter as a function of difficulty $\lambda_1 /\lambda_0$,
varying the cost of blocking NON-SPIT $c_1=k\cdot c_0$, and varying the number of calls $N=500$ or $N=5,000$ for 
the case that source=spit. 
Each entry consists of three values: $\beta^*$ the optimal setting of the accuracy obtained from minimizing 
Eq.~\eqref{eq:totalloss}; $nErr$ how often the SPIT filter made the wrong decision (out of 100,000 independent trials);
$T$ the average stopping time (number of calls). $\alpha^*$ has a constant value of 0.0001 (which is the lower boundary
 value used in optimization) in this scenario and is therefore omitted.}
\label{tab:2}
\end{table*}

\section{Network Operator's Perspective}
\label{seq:real world}
Having so far described our SPIT filter from a purely theoretical point of view, we now discuss
the steps necessary to deploy it in the real world. Note that in what follows it is neither our intent
nor within the scope of the paper to describe in detail the architecture of a fully functional
SPIT prevention system.

The section is structured as follows. First we will sketch how the SPIT filter, which should more
appropriately be seen as a SPIT detector, could be integrated into a larger SPIT prevention
system as one building block among many others. We will make suggestions on how the problem-dependent
parts of the SPRT can be instantiated by specifying:
\begin{itemize}
\item {\em sources:} how to map calls to sources such that the pure source assumption is fulfilled
\item {\em features:} what call features to use such that SPIT and NON-SPIT calls are well presented
\item {\em actions:} what action to take if the SPRT indicates a source is likely to send out SPIT
\end{itemize} 
We will then explain how the distribution of the features that discriminate SPIT from NON-SPIT can be
learned from labeled data by first assuming that the distribution is of a certain parameterized form
and then estimating these parameters from the data via maximum likelihood. In the second part of the
section we use data extracted from a large database of real-world voice calls and demonstrate 
empirically that the performance of the SPIT filter under real-world conditions with {\em a priori}
unknown distribution is also very good.

\begin{figure}
\begin{center}
\includegraphics[width=15cm]{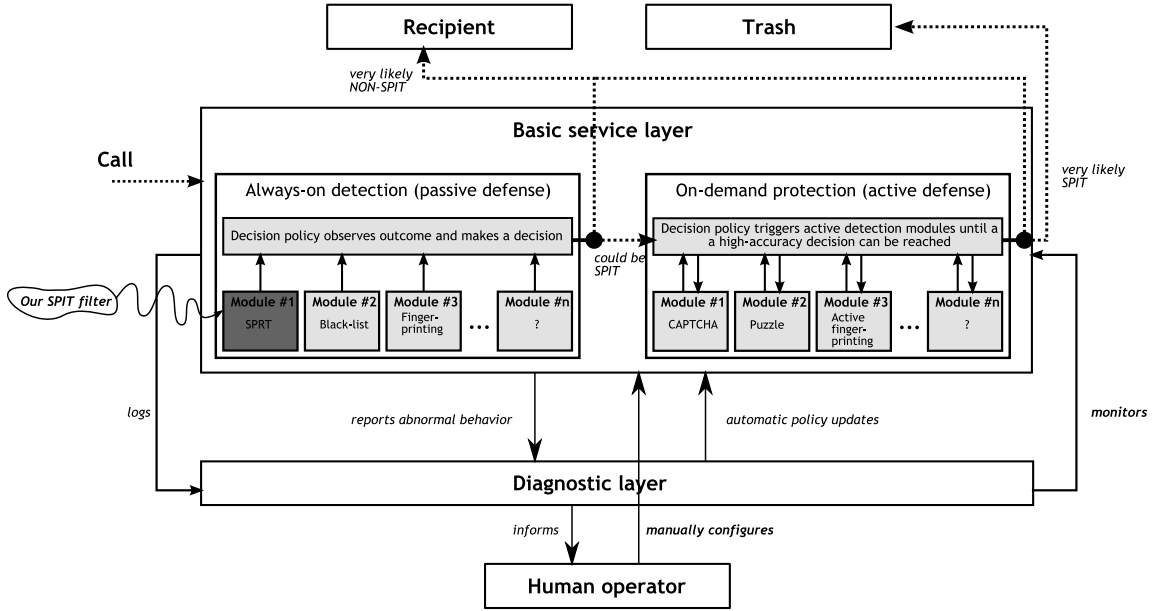}
\end{center}
\caption{A hierarchical and modular SPIT prevention system. The SPRT-based SPIT filter we propose in this paper is 
one particular module in the always-on component of the basic service layer.}
\label{fig:4}
\end{figure}

\subsection{Integration into a SPIT Prevention System}
\label{seq:how to apply}
For a network operator, a SPIT prevention system such as the one we propose and sketch in Figure~\ref{fig:4}
must allow both to maintain and guarantee an acceptable level of service under adverse operating conditions
and have a low maintenance cost. To achieve this goal, we adopt the overall strategy presented in 
\cite{resumenet-paper} which proposes a hierarchical system consisting of two modular layers: a basic
service layer and a diagnostic layer.
The basic service layer manages and processes call requests and as a whole serves to protect against 
attacks in VoIP networks -- among which SPIT is just regarded as one particular threat. For the 
prevention of SPIT the basic layer is made up of two subcomponents (a conceptually similar setup 
was also proposed for SEAL \cite{schlegel06}): {\em always-on detection} and {\em on-demand protection}.

Always-on detection consists of passive modules which essentially extract and make use of information
which is ``already there'' and thus have zero or very low computational costs. On the other hand,
these modules are only weak detectors in that they are successful only under restrictive conditions.
If the always-on detection component cannot establish with high certainty that a call is NON-SPIT,
in which case it would be allowed to pass through unharmed, the call is internally forwarded to the
on-demand protection component.

The on-demand protection component consists of active modules which require additional processing
and can have medium to high computational costs; e.g., digit-based audio CAPTCHAs 
\cite{dtmf-captcha-for-SPIT} or hidden Turing tests and computational puzzles \cite{quitt:icc}.
These modules are meant to protect the network with high accuracy. However, because of the 
cost involved (they interfere with natural communication, consume and block resources,
and may to some degree annoy human callers), they are triggered only individually and on demand.
An intelligent decision policy controls the precise way a call gets probed by the various
security tests, such that resource consumption is minimized, until a final decision SPIT
or NON-SPIT can be made with high certainty.

\subsubsection{Deployment as outbound SPIT filter}
Logically, the SPIT filter we propose would be located within the always-on component.
Physically, the SPIT filter would be located at the proxy servers which form the 
gateway between one's own network and the outside world. The SPIT filter will then act as
an {\em outbound} filter: it will perform self-moderation of outgoing calls and
unveil the presence of a SPIT botnet within one's own network before other networks
take countermeasures. Previous experience with e-mail has shown that outbound filters 
are critical to keep control over one's traffic. They ensure that the whole network's address 
space will not end up on a black list as soon as a single of its child systems becomes enrolled 
into a botnet.

\subsubsection{Defining sources}
For an outbound filter, the definition of {\em sources} -- that is, the mapping of individual call
requests to the appropriate state slot in our filter -- is straightforward since sources correspond 
to registered users and/or customers. 
The amount of information required per source (one additional number) and the number of potential sources 
itself is low enough to accommodate most operator's need without having to rely on aggregation.

\subsubsection{Defining features}
The choice of which call features to use in our SPIT filter is crucial for its performance.
While the SPIT filter is theoretically guaranteed to work with any choice of single feature
or combination of features (under which the distributions for SPIT and NON-SPIT are non-identical),
for practical reasons features with the following properties are highly desired:
\begin{itemize}
\item good separation of the distributions $\pspit$ and $\pnonspit$ as quantified by the Kullback-Leibler
information numbers $\kappa_1,\kappa_0$ from Eqs.~\eqref{eq:k0}-\eqref{eq:k1}. This ensures that the
filter will be able to stop a source from sending out further SPIT as quickly as possible.
\item hard to manipulate for spitters
\item availability of data, e.g., from old logfiles
\item easy access to the feature during runtime, meaning that the feature should be easily observable
during normal operation without requiring extra machinery.
\end{itemize} 

We believe that in this regard a good choice of features in SPIT detection are features
which capture the \emph{reaction of users} to SPIT rather than features that capture 
the technical properties of SPIT bots. Indeed, a SPIT call is:
(1) undesirable and has likely shorter duration, as the call would be hanged up by the callee 
with higher probability; (2) likely to be playing back a pre-recorded message such that 
double-talk\footnote{Double-talk means that caller and callee talk at the same time. As 
is described in \cite{wu08_voip}, this can be computed directly from the packet header information
and does not require expensive processing of the voice stream.} 
may occur; (3) unexpected with possibly longer ringing time and a higher rate of 
unanswered/\-refused calls; (4) automated with likely shorter time-to-speech and fewer pause
during the call. Although these features are more likely to be affected by cultural or social habits, 
they are much harder to manipulate for a spitter than features such as inter-arrival time or port number. 
They are also less likely to be affected by the technical characteristics of one specific botnet, 
and thus could more easily take the moving nature of SPIT attacks into account.

In this paper, we argue that {\em call duration} might be a good feature (also because it simplifies
calculation).

Of course, other choices of features are also possible. In fact, the theoretical framework in 
Section~3 allows one to do feature selection. In practice, one would thus start by identifying
a set of all possible candidate features. Given data, one would then compute the $\kappa_0$ 
and $\kappa_1$ Kullback-Leibler information number either from parametric density estimation
(as shown below in Section~\ref{sec:learningthedistribution}), or in more complicated cases
from non-parametric density estimation such as, e.g., histograms. Knowing the respective
$\kappa_1,\kappa_0$ allows one to rank the subsets and ultimately to select the 
features that achieve minimum expected regret, as the worst-case false positive and false negative
rates can be explicitly computed using the equations presented in section \ref{seq:expectedloss}.

\subsubsection{Defining decisions}
Finally we have to talk about the actions the SPIT filter can take. In Section~3 we have assumed that
once a source has been identified as a SPIT bot, all subsequent calls are to be blocked. And
conversely, once a source has been identified as a regular user, all subsequent calls are 
to be allowed through. It is clear that in practice this decision rule alone will not be
sufficient. However, recall from the beginning of this section that our SPIT filter is meant 
to be only one particular module in a larger SPIT prevention system (see Figure~\ref{fig:4}).
Thus the outcome of the SPRT should be seen as another feature by itself, based on which
a higher-level decision-making policy would then act. (The specific details of this high-level
decision-making policy are outside the scope of the paper.)

\subsection{Example: learning the distribution from labeled data}
\label{sec:learningthedistribution}
Maximum likelihood estimation is a simple way to learn the distribution from labeled data. Assume we are 
given $n$ calls either all labeled as SPIT or all labeled as NON-SPIT (without loss of generality we 
assume they are all SPIT). To estimate the distribution $\pspit$ necessary to perform the SPRT, we 
proceed as follows. First, we extract the feature representation from each call, yielding
$x_1,\ldots,x_n$. Next, we make an assumption about the form of $\pspit$; for example, assume $x_i$ 
is the call duration and we believe that an exponential distribution with (unknown) parameter $\lambda$ 
would describe the data well. To find the parameter $\lambda$ that best explains the data (under the 
assumption that the data is i.i.d. drawn from an exponential distribution) we then consider the likelihood
of the data as function of $\lambda$ and maximize it (or equivalently, its logarithm):
\begin{align}
\max_{\lambda>0} \, \mathcal{L} (\lambda) & :=\log p(x_1,\ldots,x_n|\lambda) \\
& =\log\Bigl(\prod_{i=1}^n p(x_i|\lambda)\Bigr)=\sum_{i=1}^n \log \bigl( \lambda \exp(-\lambda x_i)\bigr) \nonumber \\
& = n \log(\lambda) - \lambda \sum_{i=1}^n x_i.
\label{eq:loglike}
\end{align}
The best parameter $\lambda_{\text{ML}}$ is then found by equating the derivative of $\mathcal L(\lambda)$ with 
zero, yielding $\frac{n}{\lambda}-\sum x_i=0$, or
\begin{equation}
\lambda_{\text{ML}}=\frac{n}{\sum_{i=1}^n x_i}.
\label{eq:ml estimate}
\end{equation} 
To run the SPIT filter we would thus take $p(x|\text{SPIT}):=\lambda_{\text{ML}}\exp(-\lambda_{\text{ML}}x)$ 
in Eq.~\eqref{eq:expdist}.

\subsection{Evaluation with real world data}
\label{seq:real world experiment}
As said above, in the real world we cannot assume that we know the generating distributions $\pspit$ and $\pnonspit$. 
Instead we have to build a reasonable estimate for the distributions from labeled data. The natural question we have to answer
then is: what happens if the learned distribution used in the SPIT filter does not exactly match the true but unknown distribution
generating the data we observe (remember, the case where they do match was examined in Section~\ref{seq:experiment1}). 

To examine this point with real-world data, we used call data from 106 subjects collected from mobile phones over
several months by the MIT Media Lab and made publicly available\footnote{{\tt http://reality.media.mit.edu/download.php}} in \cite{eagle09realitydata}. The dataset gives detailed information for each call and comprises about 100,000 regular voice calls. 
Ideally we would have liked to perform the evaluation based on real-world data for both SPIT and NON-SPIT. Unfortunately,
this dataset only contains information about regular calls and not SPIT---and at the time of writing, no other such dataset
for SPIT is publicly available.

In the following we will take again call duration as feature for our filter to discriminate SPIT from NON-SPIT.
To obtain SPIT calls from the dataset, we proceed as follows. The dataset is artificially divided into two smaller
datasets: one that corresponds to SPIT and one that corresponds to NON-SPIT. The set of SPIT calls is obtained by
taking 20\% of all calls whose call duration is $<$80 seconds, the remaining calls are assigned to the set of NON-SPIT
calls.  Each time a call
is generated, we randomly select a call from the SPIT or NON-SPIT dataset, extract its call duration and forward it 
to the SPIT filter. As this time we are dealing with real-world data, we do not know the true generating distributions 
$\pspit, \pnonspit$. Instead we fit, as described in Section~\ref{sec:learningthedistribution}, an exponential 
distribution to the datasets we have built and use the learned distributions as surrogate\footnote{Note again that the 
true generating distribution is likely not exponential. Thus the theoretical bounds we derived in Section~\ref{seq:theory} 
do not directly apply. However, if the true distribution is ``close'' to an exponential, then we can expect that the 
result obtained from using only the learned distribution will also be ``close''. The experiments in this section will 
confirm that this is indeed the case.} for the unknown distribution $\pspit, \pnonspit$ in the SPIT filter 
(i.e., $\kappa_1,\kappa_0$ is calculated for the learned distributions which in our case have mean $30.23$ seconds 
for SPIT and $129.64$ seconds for NON-SPIT).

Our general experimental setup is the same as before in Section~\ref{seq:experiment1}.
We consider three scenarios: (1) scenario one, where SPIT is generated from a model and NON-SPIT is generated from the data; 
(2) scenario two, where SPIT is generated from data and NON-SPIT is generated from a model; (3) scenario three, where SPIT is
generated from data and NON-SPIT is generated from data. Here ``generated from model'' means that the generating distribution 
is exponential with known mean (thus the distribution in the SPIT filter matches the data generating distribution); 
``generated from data'' means that we use the real-world data described above (thus the distribution in the SPIT filter does not 
match the data generating distribution).

We performed 10,000,000 independent runs of the SPIT filter for each of the cases source=SPIT and source=NON-SPIT, and 
settings of the accuracy parameters $\alpha,\beta$. For each setting, Table~\ref{table:3} shows the number of times 
the SPIT filter made the wrong decision and how many calls the filter needed to observe to arrive at this decision. 
For example, when both SPIT and NON-SPIT are generated from real-world data (column 3) and with $\alpha,\beta$ set to
$10^{-3}$, the empirical error rate for NON-SPIT is $6.96\e{-2}$ (meaning that $6.96\%$ of regular calls are wrongly
identified as SPIT), while the empirical error rate for SPIT is $0$ (meaning that $0\%$ of SPIT calls are wrongly 
identified as NON-SPIT). The average number of calls the SPIT filter had to let through to arrive at this decision 
was $6.64$ and $8.18$, respectively. Note that while the error rate for NON-SPIT seems rather high (and is higher than
the expected error for this setting accuracy), we should keep in mind that in our experiment it occurs under unfavorable 
conditions (an intentional mismatch between the true model and the distribution generating the data). 

In practice, one would use more sophisticated (and more accurate) methods to estimate the distributions from data 
and, since the SPRT filter would ideally be just one component in the larger SPIT prevention framework and not be 
alone responsible for making the decision of whether to accept or reject the call, also allow higher tolerance 
thresholds for the error (which are automatically adjusted as described in Section~\ref{seq:expectedloss} 
by having a human operator define the cost of making an error appropriately).

\begin{table*}[t]
\begin{center}
{\scriptsize
\begin{tabular}{|l||c|c||c|c||c|c|}
\hline
 &  \multicolumn{2}{|c||}{SPIT=model, NON-SPIT=data} 
    &  \multicolumn{2}{|c||}{SPIT=data, NON-SPIT=model} 
    & \multicolumn{2}{|c|}{SPIT=data, NON-SPIT=data}\\
\hline
\hline
Source=SPIT  & & & & & & \\
\hline
  $\alpha,\beta$ & Error & Stopping time & Error & Stopping time & Error & Stopping time \\
\hline
                      1\e{-6}   & 0          & 15.68   
                                & 0          & 20.74 
                                & 0          & 15.67 \\
                      1\e{-5}   & 0          & 13.18   
                                & 0          & 17.48 
                                & 0          & 13.18 \\
                       1\e{-4}  & 1.10\e{-5} & 10.69   
                                & 0          & 14.07 
                                & 0          & 10.66 \\
                       1\e{-3}  & 1.13\e{-4} & 8.19   
                                & 0          & 10.73 
                                & 0          & 8.18 \\
                       1\e{-2}  & 1.43\e{-3} & 5.67   
                                & 0          & 7.38 
                                & 0          & 5.66 \\
                       1\e{-1}  & 1.74\e{-2} & 2.98   
                                & 1.10\e{-6} & 3.89 
                                & 0          & 3.05 \\
\hline
\hline
Source=NON-SPIT  & & & & & & \\
\hline
  $\alpha,\beta$ & Error & Stopping time & Error & Stopping time & Error & Stopping time \\
\hline
                      1\e{-6}   & 6.17\e{-3} & 10.10   
                                & 1.10\e{-6} & 9.30 
                                & 6.16\e{-3} & 10.11 \\
                      1\e{-5}   & 1.38\e{-2} & 9.10   
                                & 4.00\e{-6} & 8.05 
                                & 1.39\e{-2} & 9.12 \\
                      1\e{-4}   & 3.13\e{-2} & 7.96   
                                & 6.20\e{-5} & 6.79 
                                & 3.15\e{-2} & 8.00 \\
                      1\e{-3}   & 6.95\e{-2} & 6.64   
                                & 6.52\e{-4} & 5.54 
                                & 6.96\e{-2} & 6.64 \\
                      1\e{-2}   & 1.53\e{-1} & 4.96   
                                & 6.31\e{-3} & 4.25 
                                & 1.54\e{-1} & 4.97 \\
                      1\e{-1}   & 3.40\e{-1} & 2.79   
                                & 6.88\e{-2} & 2.68 
                                & 3.40\e{-1} & 2.79 \\
\hline
\end{tabular}
}
\end{center}
\caption{Results of running the SPIT filter on real-world call data. The column ``Error'' gives the empirical error as the 
average number of times the SPIT filter ended up making the wrong decision (as fraction over 10,000,000 independent runs). 
The column ``Stopping time'' gives the average number of calls necessary for the SPIT filter to arrive at a decision.}
\label{table:3}
\end{table*}

\section{Summary}
\label{seq:discussion}

In this paper, we presented the first theoretical approach to SPIT filtering that is based
on a rigorous mathematical formulation of the underlying problem and,
in consequence, allows one to derive performance guarantees in terms of worst 
case cumulative misclassification cost (the expected loss) and thus, on the number of samples 
that are required to establish with the required level of confidence that a source 
is indeed a spitter. The method is optimal under the assumption of knowing the generating 
distributions, does not rely on manual tuning and tweaking of parameters, and is 
computationally simple and scalable.
These are desirable features that make it a component of choice in a larger, autonomic framework.

Moreover, we have outlined the procedure that needs to be followed 
to apply this SPIT filter as an \emph{outbound} filter in a realistic SPIT prevention 
system, including which potential call features to use and how the best feature could 
be found from the candidates via automated feature selection. In particular, we have 
sketched how the generating distributions can be learned from data. The difficulty of 
the problem of successfully detecting SPIT is then only related to how similar/dissimilar 
the generating distributions are. This difficulty can be quantitatively expressed in terms
of the Kullback-Leibler information numbers $\kappa_1,\kappa_0$---which in turn can be 
calculated analytically or approximately from the learned distributions. Taken together 
this means that the worst case performance of the SPIT filter can be computed in real-world
operation (and can thus be potentially used to tune the other hyperparameters of the 
whole system).

Our experimental evaluation, both on data synthetically 
generated and on data extracted from real-world call data, verifies that our approach is   
feasible, efficient (``efficient'' meaning that only very few calls need to be observed from a
source to identify SPIT), and able to produce highly accurate results even when the 
generating distribution is not {\em a priori} specified but inferred from data.

\section*{Acknowledgements}
{\small Damien Ernst (Research Associate) and Sylvain Martin (Post-Doctoral Researcher) acknowledge the 
financial support of the Belgian Fund of Scientific Research (FNRS).This work is also partially funded by 
{EU} project {R}esume{N}et, FP7--224619.}

\bibliographystyle{plain}

\begin{thebibliography}{10}

\bibitem{Duffy94statisticalanalysis}
D.~E. Duffy, A.~A. Mcintosh, M.~Rosenstein, and W.~Willinger.
\newblock Statistical analysis of ccsn/ss7 traffic data from working ccs
  subnetworks.
\newblock {\em IEEE JSAC}, 1994.

\bibitem{eagle09realitydata}
N.~Eagle, A.~Pentland, and D.~Lazer.
\newblock Inferring social network structure using mobile phone data.
\newblock {\em Proceedings of the National Academy of Sciences (PNAS)},
  106(36):15274--15278, 2009.

\bibitem{B08}
D.~Geneiatakis and C.~Lambrinoudakis.
\newblock A lightweight protection mechanism against signaling attacks in a
  sip-based voip environment.
\newblock {\em Telecommunication Systems}, 36(4):153--159, 2008.

\bibitem{Kolan07}
P.~Kolan and R.~Dantu.
\newblock Socio-technical defense against voice spamming.
\newblock In {\em ACM Transactions on Autonomous and Adaptive Systems (TAAS)},
  2007.

\bibitem{nassar10}
M.~Nassar, O.~Dabbebi, R.~Badonnel, and O.~Festor.
\newblock Risk management in voip infrastructure using support vector machines.
\newblock In {\em International conference on Network and Service Management
  (CNSM'10)}, pages 48--55, 2010.

\bibitem{nassar-SVM}
M.~Nassar, R.~State, and O.~Festor.
\newblock Monitoring sip traffic using support vector machines.
\newblock In {\em Proceedings of the 11th international symposium on Recent
  Advances in Intrusion Detection}, RAID '08, pages 311--330, Berlin,
  Heidelberg, 2008. Springer-Verlag.

\bibitem{quitt:icc}
J.~Quittek, S.~Niccolini, S.~Tartarelli, M.~Stiemerling, M.~Brunner, and
  T.~Ewald.
\newblock Detecting {SPIT} calls by checking human communication patterns.
\newblock In {\em {IEEE} International Conference on Communications (ICC
  2007)}, June 2007.

\bibitem{Self-learning08}
K.~Rieck, S.~Wahl, P.~Laskov, P.~Domschitz, and K.-R. M\"uller.
\newblock Self-learning system for detection of anomalous sip messages.
\newblock In {\em Principles, Systems and Applications of IP
  Telecommunications, 2nd International Conference, IPTComm (2008)}, pages
  90--106, 2008.

\bibitem{Robbins52}
H.~Robbins.
\newblock Some aspects of the sequential design of experiments.
\newblock {\em Bulletin of American Mathematical Society}, 58:527--535, 1952.

\bibitem{schlegel06}
R.~Schlegel, S.~Niccolini, S.~Tartarelli, and M.~Brunner.
\newblock {SPIT} prevention framework.
\newblock In {\em IEEE GLOBECOM'06}, pages 1--6, 2006.

\bibitem{Shin06}
D.~Shin, J.~Ahn, and C.~Shim.
\newblock Progressive multi gray-leveling: a voice spam protection algorithm.
\newblock {\em IEEE Network}, 20:18--24, 2006.

\bibitem{spider-final-report08}
Y.~Soupionis, G.~Marias, S.~Ehlert, Y.~Rebahi, S.~Dritsas, M.~Theoharidou,
  G.~Tountas, D.~Gritzalis, A.~Bergmann, T.~Golubenco, and M.~Hoffmann.
\newblock {SPAM over Internet telephony Detection sERvice} final report.
\newblock {\tt
  \verb+http://projectspider.org/documents/Spider_D4.2_public.pdf+}, Sep 2008.

\bibitem{dtmf-captcha-for-SPIT}
Y.~Soupionis, G.~Tountas, and D.~Gritzalis.
\newblock Audio {CAPTCHA} for {SIP}-based {VoIP}.
\newblock In {\em Emerging Challenges for Security, Privacy and Trust}, volume
  297 of {\em IFIP Advances in Information and Communication Technology}, pages
  25--38, 2009.

\bibitem{resumenet-paper}
J.~Sterbenz, D.~Hutchison, E.~K. \c{C}etinkaya, A.~Jabbar, J.~P. Rohrer,
  M.~Sch\"{o}ller, and P.~Smith.
\newblock Resilience and survivability in communication networks: Strategies,
  principles, and survey of disciplines.
\newblock {\em Computer Networks}, 54:1245--1265, June 2010.

\bibitem{Wald43}
A.~Wald.
\newblock Sequential tests of statistical hypotheses.
\newblock {\em Annals of Mathematical Statistics}, 16:117--186, 1945.

\bibitem{WaldWolf48}
A.~Wald and J.~Wolfowitz.
\newblock Optimum character of the sequential probability test.
\newblock {\em Annals of Mathematical Statistics}, 19:326--339, 1948.

\bibitem{wu08_voip}
C.-C. Wu, K.-T. Chen, Y.-C. Chang, and C.-L. Lei.
\newblock Detecting voip traffic based on human conversation patterns.
\newblock In Henning Schulzrinne, Radu State, and Saverio Niccolini, editors,
  {\em Principles, Systems and Applications of IP Telecommunications. Services
  and Security for Next Generation Networks}, volume 5310 of {\em Lecture Notes
  in Computer Science}, pages 280--295. Springer Berlin / Heidelberg, 2008.

\bibitem{wu09}
Y.-S. Wu, S.~Bagchi, N.~Singh, and R.~Wita.
\newblock Spam detection in voice-over-ip calls through semi-supervised
  clustering.
\newblock In {\em Proceedings of the 2009 Dependable Systems Networks}, pages
  307--316, 2009.

\bibitem{C06}
H.~Yan, K.~Sripanidkulchai, H.~Zhang, Z.-Y. Shae, and D.~Saha.
\newblock Incorporating active fingerprinting into spit prevention systems.
\newblock In {\em Third annual security workshop (VSW'06)}, 2006.

\bibitem{A07}
G.~Zhang, S.~Ehlert, T.~Magedanz, and D.~Sisalem.
\newblock Denial of service attack and prevention on sip voip infrastructures
  using dns flooding.
\newblock In {\em Principles, Systems and Applications of IP
  Telecommunications, 1st International Conference, IPTComm (2007)}, 2007.

\bibitem{AA10}
G.~Zhang, S.~Fischer-H\"ubner, and S.~Ehlert.
\newblock Blocking attacks on sip voip proxies caused by external processing.
\newblock {\em Telecommunication Systems}, 45(1):61--76, 2010.

\end{thebibliography}

\end{document}